# Boosting hydrogen and methane formation on a high-entropy photocatalyst by integrating atomic $d^0/d^{10}$ electronic junctions and microscopic P/N heterojunctions

Ho Truong Nam Hai[a,b], Jacqueline Hidalgo-Jiménez[a,b] and Kaveh Edalati[a,b,c,]*

[a] WPI, International Institute for Carbon Neutral Energy Research (WPI-I2CNER), Kyushu University, Fukuoka 819-0395, Japan
[b] Department of Automotive Science, Graduate School of Integrated Frontier Sciences, Kyushu University, Fukuoka 819-0395, Japan
[c] Mitsui Chemicals, Inc. - Carbon Neutral Research Center (MCI-CNRC), Kyushu University, Fukuoka, 819-0395, Japan

The formation of green energy carriers such as hydrogen ($H_2$) and methane ($CH_4$) via photocatalytic processes provides a clean method for addressing environmental and energy issues. To achieve highly efficient photocatalysts for $H_2$ and $CH_4$ generation, the present work introduces the P/N heterojunctions in a high-entropy oxide (HEO) with $d^0/d^{10}$ electronic junctions. The study uses CuO as a P-type semiconductor and the HEO containing $d^0$ (Ti, Zr, Nb, Ta) and $d^{10}$ (Zn) cations as an N-type semiconductor. The material exhibits improvements in optical properties, such as light absorption, charge mobility and reduced electron-hole recombination. The integration of two concepts, atomic-scale $d^0/d^{10}$ electronic junctions and micro-scale P/N heterojunctions, leads to enhanced $H_2$ and $CH_4$ production. Particularly after the partial removal of vacancies in the heterojunction, $H_2$ production from photocatalytic water splitting reaches 0.71 mmol/g.h, and $CH_4$ evolution from $CO_2$ conversion reaches 2.40 μmol/g.h with 72% selectivity for methanation. The integrated strategy of this study has a high potential in developing active heterostructured catalysts for clean fuel production.

*Corresponding author: Kaveh Edalati (E-mail: kaveh.edalati@kyudai.jp; Phone: +81 92 802 6744)

# 1. Introduction

In recent decades, humans have been actively searching for new energy carriers to ensure energy security that meets the rapidly growing demands driven by scientific and technological advancements, instead of solely relying on fossil fuels [1,2]. Moreover, society faces prominent environmental challenges from fossil fuel consumption, one of them being the significant increase in carbon dioxide ($CO_2$) emissions [3,4]. The $CO_2$ gas is the main cause of the greenhouse effect, which contributes to extreme weather events [5,6]. Therefore, finding methods to convert $CO_2$ into fuels or using hydrogen ($H_2$), as a zero $CO_2$-emitting gas, presents a major challenge in energy and environmental issues. Photocatalysis is a green and sustainable technology that can be utilized for such chemical reactions for $CO_2$ conversion and $H_2$ production by harnessing unlimited energy from the sun using semiconducting catalysts with appropriate bandgaps [7,8].

A typical photocatalytic process consists of three main stages [7,9–11]. In the first stage, the catalyst absorbs sunlight and the electrons oscillate [7]. When the energy level exceeds the bandgap of the catalyst, electrons move to the conduction band minimum (CBM) from the valence band maximum (VBM), and simultaneously, holes are formed [9]. Subsequently, the charge carriers (electrons and holes) may recombine or move to the catalyst surface [11]. Finally, in the interface of the catalyst surface with the solution, these electrons and holes directly or indirectly participate in chemical reactions to form products such as $CH_4$ from $CO_2$ and $H_2$ from water [10,12]. This catalytic process does not produce harmful products and exclusively uses solar radiation as an energy source. However, the majority of catalysts have a large bandgap and accordingly, low light absorption [13–15]. Besides, the fast electron and hole recombination, excited by light absorption, reduces the efficiency of the photocatalytic processes for producing desired products [16].

As a solution to enhance photocatalytic activity by suppressing the charge carrier recombination, designing the P/N heterojunction structure can be considered an effective approach [16,17]. The P/N heterojunction is generated through the combination of two kinds of semiconductors in the catalyst structure, with holes dominated in the P layer or P phase, while electrons accumulate in the N layer or N phase [17]. P/N heterojunctions are divided into three main groups of broken gap, straddling gap and staggered gap based on the bandgap of each P- and N-type semiconductor [16–22]. In heterojunctions with a staggered gap structure as the most effective type for photocatalysis, also referred to as a type II heterojunction [17,20], photoinduced electrons move from one semiconducting phase having a higher conduction band (CB) edge to another semiconductor having a lower CB edge, contrasted with the photoinduced holes moving from the lower valence band (VB) edge to the higher VB edge [17,20]. This arrangement helps to enhance charge transfer at interfaces of semiconductors and significantly reduce the recombination of excited electrons and holes [20].

Another strategy to suppress recombination is to include both electron donor and acceptor cations having $d^0$ and $d^{10}$ electronic configurations in the atomic structure of a photocatalyst [23]. While the P/N junction strategy links two different phases, the mixed $d^0$ and $d^{10}$ electronic configuration, recently realized in high-entropy oxides (HEOs), links different types of atoms in a single phase. HEOs are oxides of a minimum of five cations in a crystal lattice, having high structural stability and significant chemical composition tuning capability [24–29]. Additionally, HEOs are characterized by a strained structure and the formation of oxygen vacancies [30], a key parameter that allows HEOs to function as an N-type catalyst. Due to these features, HEOs have been studied in various catalytic processes, such as $CO_2$ conversion [24], lithium storage [26], wastewater remediation [27], $H_2$ production [28] and oxygen ($O_2$) production [29]. Following the introduction of high-entropy photocatalysts in 2020 [31], they have been used for various photocatalytic reactions such as $H_2$ production [32], $O_2$ production [33], $H_2O_2$ production [34], $CO_2$ conversion [35], dye degradation [36] and plastic waste degradation [37]. Although recent studies reported high photocatalytic performance of these materials with atomic $d^0/d^{10}$ electronic junctions [23], there have been no attempts to date to introduce P/N heterojunctions in them to further boost their activity for clean $CH_4$ and $H_2$ production.

In this study, a heterostructured catalyst with P/N heterojunction is synthesized based on an HEO with atomic $d^0/d^{10}$ electronic junctions (TiZrNbTaZnO$_{10}$) as an N-type semiconducting oxide and CuO as a P-type semiconducting oxide through the high-pressure torsion (HPT) [38] technique and subsequent annealing. Although the significance of the $d^0/d^{10}$ concept on the enhancement of photocatalytic efficiency of HEOs has been reported in a few studies [39,40], this work reports the first attempt in integrating this concept with P/N heterojunctions to enhance charge separation and photocatalytic efficiency. TiZrNbTaZnO$_{10}$ was selected as it is one of the limited available single-phase high-entropy photocatalysts with $d^0/d^{10}$ electronic junctions [23], and CuO was selected as it is one of the limited P-type semiconductors being used for photocatalysis [41]. The introduction of P/N heterojunction improves both methanation and water splitting, demonstrating the great potential of the utilized strategy for photocatalytic processes to generate clean fuels.

## 2. Materials and methods

### 2.1. Reagents

Titanium dioxide ($TiO_2$) anatase 98 %, copper (II) oxide (CuO) 99 %, zirconium dioxide ($ZrO_2$) 97 %, niobium pentoxide ($Nb_2O_5$) 99 %, tantalum pentoxide ($Ta_2O_5$) 99.9 % and zinc oxide (ZnO) 99 %, were bought from different companies, including Sigma-Aldrich, USA, Kojundo, Japan, Kanto Chemical Co. Inc., Japan. Hexachloroplatinic (IV) acid hexahydrate ($H_2PtCl_6 \cdot 6H_2O$) was purchased from Goodfellow, UK.

### 2.2. Synthesis of photocatalysts

A composite containing a P/N heterojunction was synthesized in three main steps. First, the Ti-Zr-Nb-Ta-Zn-based HEO was prepared following the method in a previous paper [23]. A powder mixture containing five binary oxides with the same atomic percent of cations was mixed in acetone for 30 minutes. After that, the mixed and dried oxides were compacted into a 0.8 mm thick and 5 mm radius disk (352 mg/disk) and further mixed by HPT (pressure 6 GPa, rotation 3 turns for two times, speed 1 rpm, temperature 298 K) followed by a calcination process (temperature 1373 K, time 24 h). The calcined material was further mixed by 3 turns of HPT and calcined again for another 24 h. In the second processing step, the HEO (50 vol%) and CuO nanopowder (50 vol%) were mixed in acetone for 30 minutes and then pressed into a disk (395 mg/disk). Subsequently, the disk was treated by HPT (pressure 6 GPa, rotation 3 turns, speed 1 rpm, temperature 298 K) to create the heterojunction structure between the HEO and CuO. In the third step, the HPT-treated composite with heterojunction was annealed for 1 h at 973 K for the removal of oxygen vacancies as a strategy to increase the activity of the heterostructured catalyst [42].

### 2.3. Characterizations

The structural features of three catalysts, including the HEO, the HPT-treated HEO/CuO and the annealed HEO/CuO, were studied through X-ray diffraction (XRD) with a copper Kα lamp and Raman spectra with a 532 nm wavelength laser. A scanning electron microscope (SEM) at 15 kV was used to investigate the microstructure and evaluate elemental distribution using energy-dispersive X-ray spectroscopy (EDS). A scanning-transmission electron microscope (STEM) at 200 kV was used for the examination of nanostructure using high-resolution images, selected area electron diffraction (SAED), high-angle annular dark-field (HAADF) micrographs and EDS. The samples for TEM were made by crushing the catalysts in ethanol and distributing them onto carbon-coated grids. X-ray photoelectron spectroscopy (XPS) was used, with an Al Kα light, to determine the oxidation states of the cations and the reduction state for oxygen, as well as to calculate the VBM of the catalysts. Additionally, a differential scanning calorimetry (DSC) device, with an air atmosphere and a heating speed of 5 K min$^{-1}$, was used to characterize oxygen vacancies. Light absorption experiments of catalysts in the 200-800 nm wavelength range were conducted using UV-Vis diffuse reflectance spectroscopy. The estimation of the bandgap value was performed based on the Kubelka-Munk approach. The charge carrier recombination was determined by photoluminescence (PL) spectra, using a 325 nm He-Cd laser. The charge carrier lifetime was examined by time-resolved photoluminescence spectroscopy using a laser source with a 340 nm wavelength and a 385 nm long-pass filter. The bi-exponential decay model was used to fit the time-resolved PL data, as shown in Eqs. 1-2 [25].

$$I(t) = A_1 e^{-t/\tau_1} + A_2 e^{-t/\tau_1} \tag{1}$$

$$\tau_{avg} = \frac{A_1 \tau_1^2 + A_2 \tau_2^2}{A_1 \tau_1 + A_2 \tau_2} \tag{2}$$

In these equations, $I(t)$ is the fluorescence intensity over time $t$ after pulsed laser excitation, $\tau_1$ and $\tau_2$ represent the fast and slow recombination components, $\tau_{avg}$ is the average carrier lifetime and $A_1$ and $A_2$ are relative component amplitudes. The stability and mobility of photo-excited electrons were also determined through photocurrent experiments in the potentiostatic amperometry mode. The sample preparation process was performed following the procedure explained elsewhere [10], where 10 mg of catalysts were crushed in ethanol and coated onto a fluorine-doped tin oxide (FTO) glass (1.7×3.3 cm$^2$), then annealed at 473 K for 2 h to form a thin film. The current intensity measurement was performed for several cycles with illumination for 60 s and chopping the light for 60 s. For photocurrent measurements, 1 M KOH electrolyte, a reference Ag/AgCl electrode, a platinum counter electrode and the catalyst-coated FTO glass as a working electrode were used.

**2.4. Catalytic experiments**

Two reactions, $CO_2$ conversion and water splitting, were examined to evaluate the performance of the catalysts. For $CO_2$ conversion, the experiment was conducted with 0.1 g of catalysts and 0.1 M $NaHCO_3$ (4.2 g in 0.5 L deionized water) in a cylindrical-shaped quartz photoreactor with a 0.858 L volume. A high-pressure mercury lamp with 400 W overall power was placed at the center of the reactor. The illumination on the photoreactor was reasonably uniform with a power intensity of 1.4 W cm$^{-2}$ (measured by an optical power meter, 843-R model, Newport Corporation, Japan). The gas of $CO_2$ was injected into the solution from an inlet pipe and flowed out through an outlet pipe at the top of the photoreactor. The outlet pipe was connected directly to two gas chromatographs (GC), in which GC #1 had a thermal conductivity detector (TCD) to identify $H_2$ and $O_2$ and GC #2 had a methanizer and a flame ionization detector (FID) to measure $CH_4$ and CO. Additionally, the reactor was connected to a water chiller and put on a stirrer to keep the temperature in the range of 291-296 K while stirring at a rate of 420 rpm. A blank experiment was carried out for the first 30 minutes, with only the catalyst present and the $CO_2$ gas continuously supplied without irradiation to confirm that the $CO_2$ conversion process did not occur without illumination.

For the water splitting process, a catalyst with a mass of 0.05 g was added to a 160 mL quartz round-bottom flask, followed by the addition of 27 mL of distilled pure water, 3 mL of methanol as a hole scavenger and 0.25 mL of 0.01 M $H_2PtCl_6 \cdot 6H_2O$ as a co-catalyst. The flask was sealed using a septum cap and continuously bubbled with argon for 10 minutes to create an inert environment. The flask was then placed in a cooling bath and stirred to keep the temperature between 291 and 296 K while stirring at a speed of 420 rpm. A 300 W Xe light source was positioned 10 cm above the flask to have a uniform light intensity on the reactants with an intensity of $1.3 \pm 0.1$ W/cm$^2$ (measured by an optical power meter, 843-R model, Newport Corporation, Japan). The sampling was performed by taking 0.5 mL of gas from the flask by a syringe and injecting it directly into the GC with a TCD detector. $H_2$ concentration was calculated from the peak area in GC data using calibration curves (e.g., $y = 256846x$ with $R^2 = 0.991$, where $x$ indicates $H_2$ amount in μmol and $y$ indicates area under $H_2$ peak). The detection limit of the TCD detector was 1 μmol, and the total error of measurements was less than 10%. In addition to the photocatalytic test, a blank experiment was performed under dark conditions to confirm that water splitting does not occur without illumination.

## 3. Results
### 3.1. Catalyst characterization

Fig. 1a depicts the crystal structure of the catalysts by XRD. Peak deconvolution and Rietveld analysis (PDXL software) suggest the presence of two monoclinic phases with space groups of Cc and P2/c with a rough mass ratio of 6:4. The monoclinic phase is the phase of both CuO (Cc group, $a = 0.4688$ nm, $b = 0.3427$ nm, $c = 0.5132$ nm; $\alpha = \gamma = 90°$, $\beta = 99.47°$) and the HEO (P2/c group, $a = 0.4738$ nm, $b = 0.5651$ nm, $c = 0.5050$ nm; $\alpha = \gamma = 90°$, $\beta = 91.14°$) [23,43], and the mass ratio is in good agreement with the designated volume ratio of 1:1. After the HPT process, the heterostructured composite still maintains the characteristic phases of the two materials without showing any phase transformation. However, most of the peaks of the two phases are broadened, which is a characteristic of the lattice distortion caused by structural defects after HPT processing [38]. Subsequently, the post-HPT annealing process leads to no phase transformations but results in a reduction in the broadening of the peaks due to thermal annihilation of crystal lattice defects. Fig. 1b illustrates the Raman spectra, measured at four different positions of the catalysts. The Raman spectrum exhibits three peak positions corresponding to the $A$g and $B$g vibrational modes in the CuO structure [44]. Meanwhile, the peaks in the Raman spectrum of HEO are difficult to observe due to their low intensity and weak sharpness, as reported in an earlier study [23]. Fig. 1 confirms that the dual-phase heterostructured composite produced by HPT and post-HPT annealing shows no phase transformation or chemical reactions. The absence of transformations is essential for photocatalytic performance because any reaction at the interphase boundary leads to the ineffectiveness of the heterojunction concept [43,45].

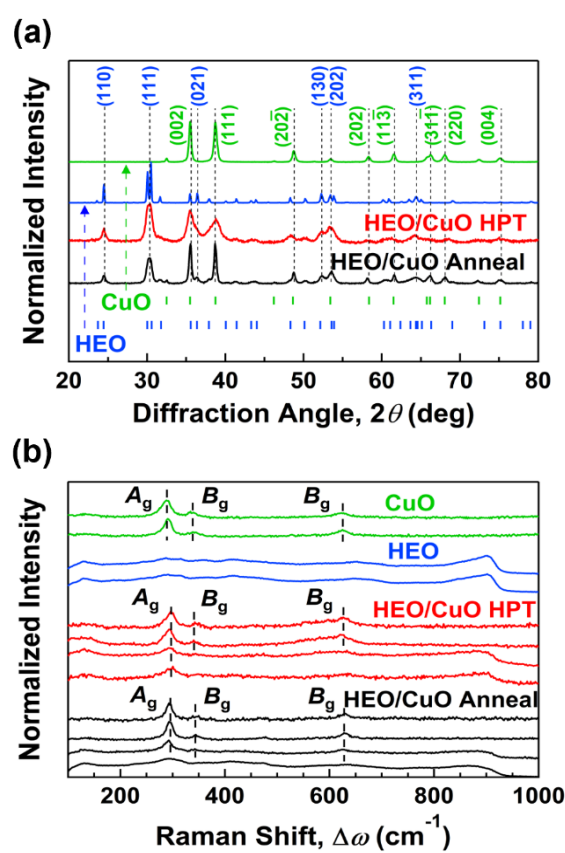

Fig. 1. Formation of heterostructured composites from high-entropy oxide without phase transformation. (a) XRD and (b) Raman profiles of CuO, HEO, HPT-processed HEO/CuO and annealed HEO/CuO.

Fig. 2 illustrates (a, c, e) SEM and (b, d, f) STEM micrographs and relevant EDS maps for (a, b) the HEO, (c, d) the HPT-treated HEO/CuO and (e, f) the annealed HEO/CuO. Fig. 2a and Fig. 2b show the highly homogeneous distribution of the cations in the HEO at the micrometer and nanometer scales. In addition, SEM-EDS analysis indicates that the HEO contained 5.4 at% Ti, 5.9 at% Zr, 5.8 at% Nb, 5.4 at% Ta, 4.2 at% Zn and 73.3 at% O, indicating the efficiency of the five-step thermomechanical oxidation synthesis to make an HEO with a composition close to the nominal composition of $TiZrNbTaZnO_{10}$. Note that the oxygen fraction is slightly overestimated due to the presence of oxygen in the carbon tape used for SEM sample preparation. For the heterostructured composites, Fig. 2c-f depicts that the HEO and CuO were mixed well at the nanometer level, leading to the presence of large fractions of interphase boundaries as heterojunction sites. Such a good mixing of phases is due to the effect of HPT on the mechanical mixing under high pressure [38].

The microstructural characterization of the catalysts is depicted in Fig. 3. TEM bright-field images in Fig. 3a, 3d and 3i confirm the presence of nanograins. SAED analysis in Fig. 3b, 3e and 3j shows the concentric ring-like shape, which is considered characteristic of nanocrystalline samples. In addition, Fig. 3b depicts the indices of the HEO while Fig. 3e and Fig. 3j show the indices of the HEO and CuO in heterojunction, which is in agreement with the XRD profiles in Fig. 1a. Fig. 3f and 3k show a higher magnification of the microstructure in the catalysts together with (c, g, h, l, m) inverse fast Fourier transform at selected regions. It confirms that the boundaries between CuO and HEO nanocrystals are formed in the HPT-processed and annealed HEO/CuO catalysts. These results demonstrate success in synthesizing heterojunctions in a catalyst based on HEOs.

Fig. 4 depicts the catalyst XPS data, in which peak fitting is based on references from the Handbook of XPS [46] and XPSPEAK41 software (background subtraction by Tougaard method, peak deconvolution by mixed Gaussian-Lorentzian method, and error margins of ±0.1–0.3 eV for peak position and 5-10 % for peak intensity). The cations in the HEO and heterostructured composites exhibit full oxidation states, including $Ti^{4+}$, $Zr^{4+}$, $Nb^{4+}$, $Ta^{5+}$, $Zn^{2+}$ and $Cu^{2+}$. Fig. 4a, 4c and 4e show a small shift to lower energy for Ti $2p_{3/2}$, Nb $3d_{5/2}$ and Zn $2p_{3/2}$ peaks in the HPT-processed HEO/CuO, compared to those in the HEO. Additionally, Fig. 4b and Fig. 4d illustrate that the peak positions for Zr $3d_{5/2}$ and Ta $4f_{7/2}$ remain almost unchanged across the three catalysts. When comparing the HPT-treated and annealed HEO/CuO catalysts, the XPS peaks for Ti $2p_{3/2}$ and Zn $2p_{3/2}$ move to lower energies following HPT treatment and return to the initial energy by annealing, but the peaks of Nb $3d_{5/2}$ stay at the same position. Moreover, a shift to lower energies is observed for Cu $2p_{3/2}$ peak after annealing (Fig. 4f). These energy changes are likely because of the generation of oxygen vacancies following the HPT treatment and their removal by the annealing treatment [47].

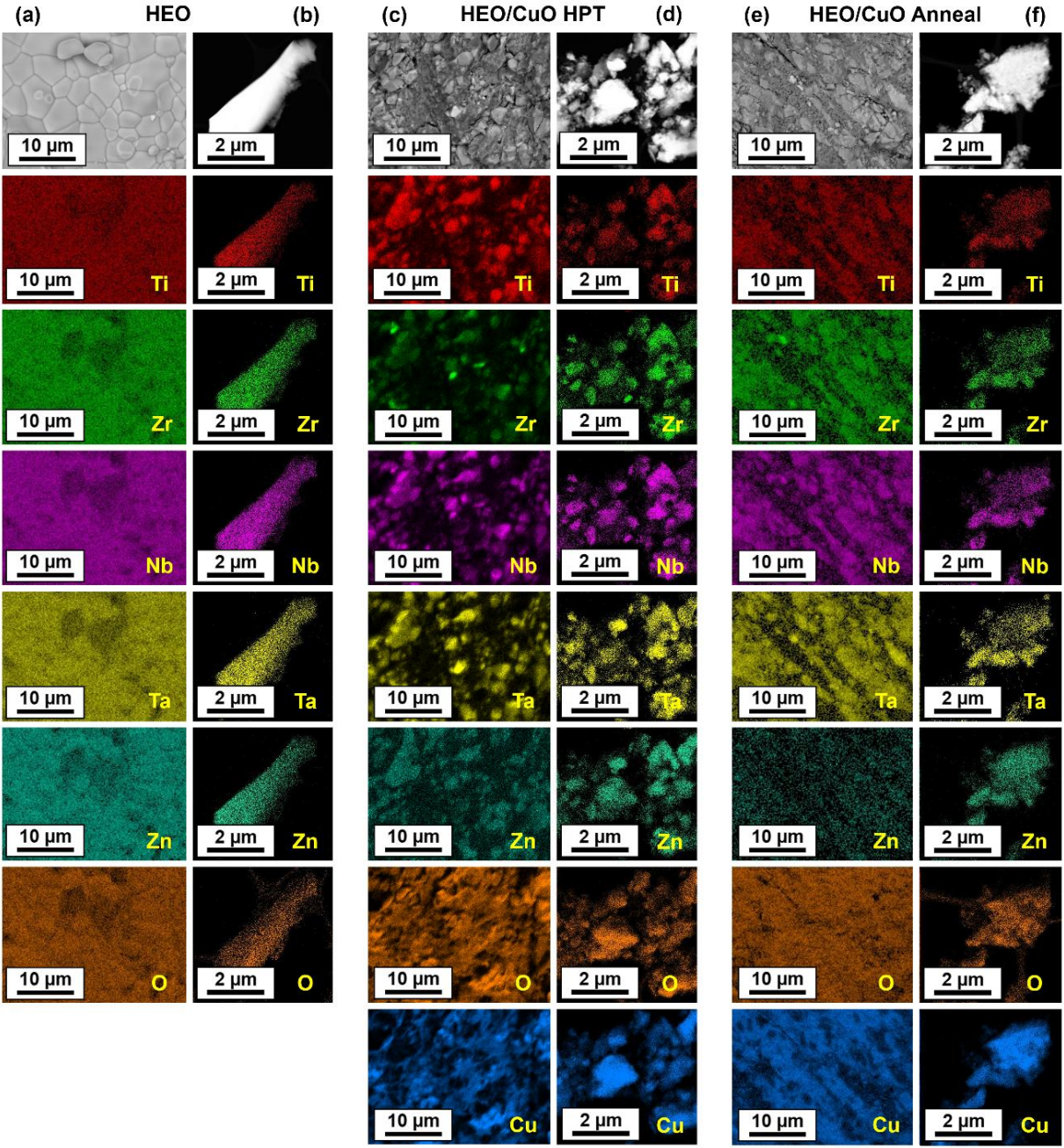

Fig. 2. Appropriate mixing of copper oxide and high-entropy oxide in micrometer and nanometer scales to form a heterostructured composite. (a, c, e) SEM and (b, d, f) STEM micrographs and relevant EDS maps for (a, b) HEO, (c, d) HPT-treated HEO/CuO and (e, f) annealed HEO/CuO.

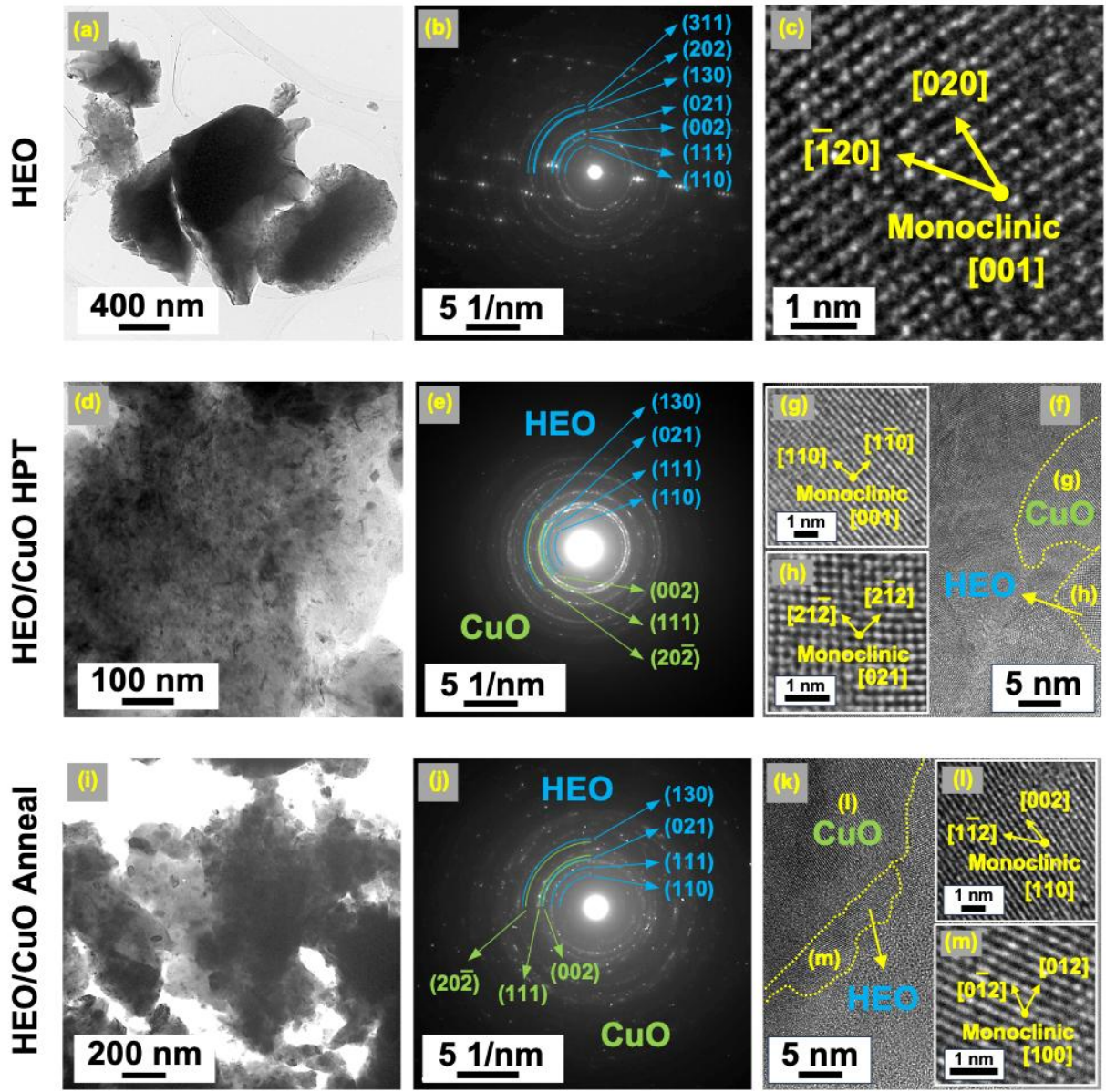

Fig. 3. Formation of interphase boundaries between high-entropy oxide and copper oxide. (a, d, i) TEM bright-field micrographs, (b, e, j) SAED patterns, (f, k) high-resolution images and (c, g, h, l, m) lattice images of (a, b, c) HEO, (d, e, f, g, h) HPT-processed HEO/CuO and (i, j, k, l, m) annealed HEO/CuO.

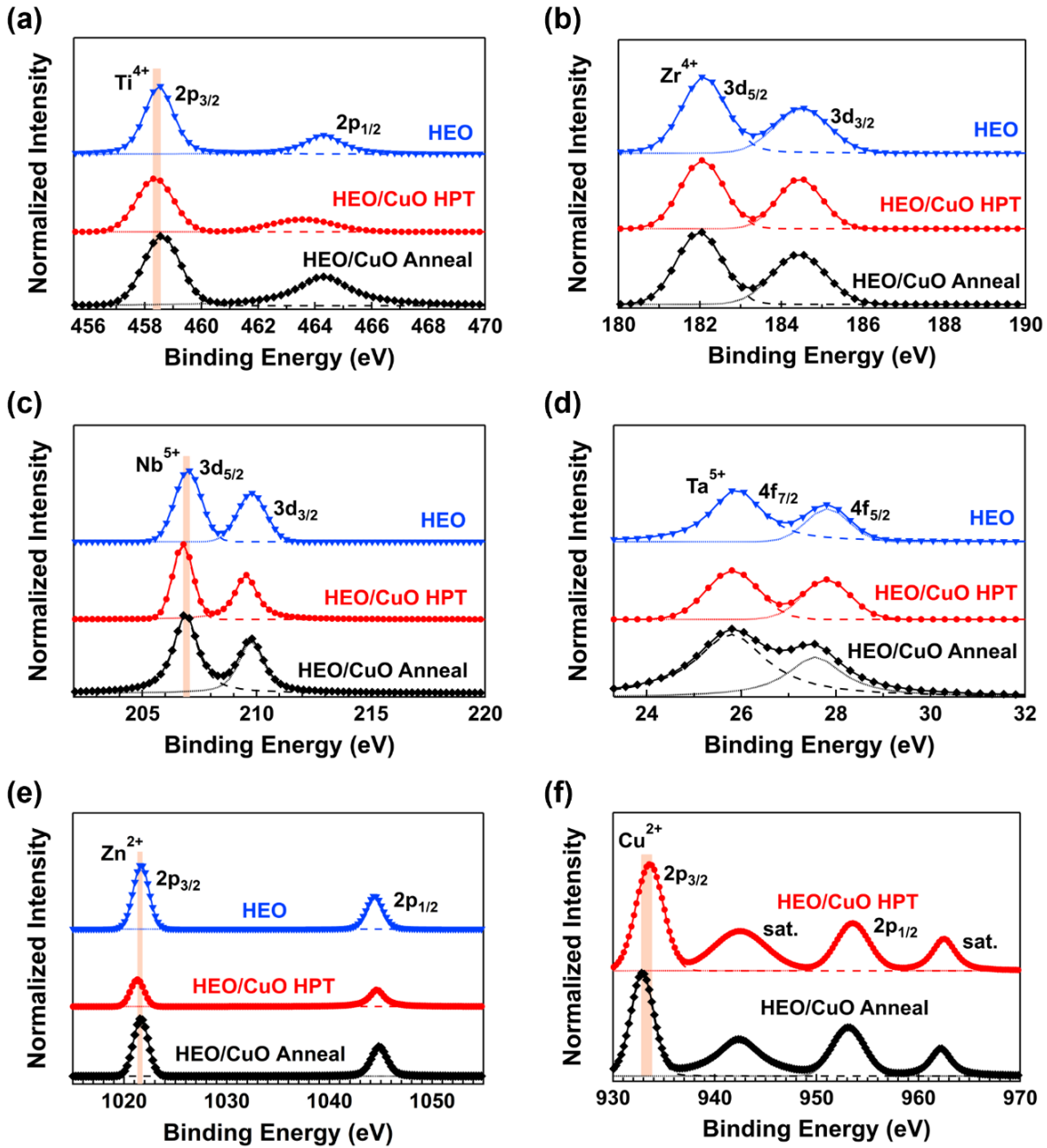

Fig. 4. Oxidation states of cations in high-entropy oxide and corresponding heterostructured composites. XPS spectra and relevant peak deconvolution of (a) titanium 2p, (b) zirconium 3d, (c) niobium 3d, (d) tantalum 4f, (e) zinc 2p and (f) copper 2p for (a-e) HEO, (a-f) HPT-processed HEO/CuO and (a-f) annealed HEO/CuO.

The generation of oxygen vacancies was studied using (a) O 1s XPS and (b) DSC, as illustrated in Fig. 5. The O 1s XPS spectrum of the HEO shows lattice oxygen $O^{2-}$ at a peak location of 530.3 eV. The lattice oxygen peak location slightly shifts to the lower energy at 529.8 in the HPT-processed HEO/CuO, which is explained by the domination of O 1s in CuO [48] on the surface of heterostructured composites after HPT. Additionally, a new peak corresponding to the OH group absorption and oxygen vacancies appears at 531.3 eV [49,50]. For the annealed HEO/CuO, the peak at 531.3 eV shows a significant decrease in intensity, suggesting that the concentration of oxygen vacancies on the surface is reduced through the annealing process. Fig. 5b presents the DSC spectra of the three catalysts achieved during heating under air. The HEO shows a continuous increase in heat flow without the appearance of any peak, but the HPT-processed HEO/CuO shows an exothermic peak at 770 K, with a total energy of 0.17 kJ/mol. Examination of the material by XRD before and after this peak confirms the absence of phase transformation, and thus, this peak should originate from the annihilation of crystal imperfections such as vacancies. For the annealed HEO/CuO, which was intentionally annealed after this peak, the peak cannot be detected in Fig. 5b. Vacancy formation energy in HEOs strongly depends on the localized configuration, but values between 0.17 eV and 4.59 eV were reported in a few publications [51,52]. If it is assumed that the 0.17 kJ/mol energy in DSC comes only from oxygen vacancies, the fraction of vacancies should be roughly in the range of 0.04 to 1%. Since vacancies with lower formation energy are more likely to form, 1% is theoretically a better estimate for the vacancy fraction. It should be noted that vacancy concentrations above the equilibrium vacancy concentration are not thermodynamically stable; however, synthesis methods like HPT can lead to supersaturation of vacancies [53–55]. The presence of such excess vacancies in HPT-processed ceramic catalysts is a particularly general feature due to low atomic diffusion and slow vacancy annihilation kinetics at ambient temperature [56,57], and thus, to remove such excess vacancies, thermal treatments like annealing are usually needed. Fig. 5 suggests that the synthesis of the heterostructured composite based on the HEO via the HPT method leads to the generation of oxygen vacancies, but such vacancy-type imperfections are subsequently eliminated through the annealing process.

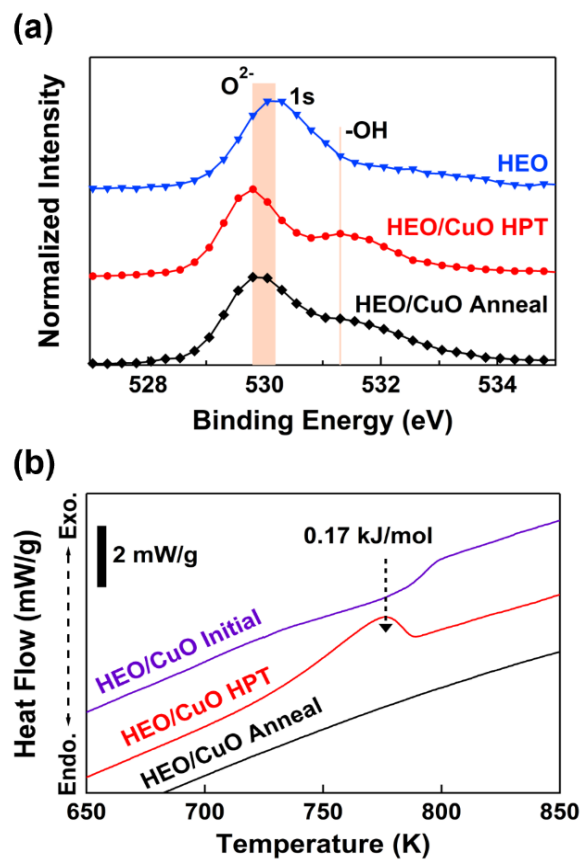

Fig. 5. Generation of oxygen vacancies by high-pressure torsion and removal of them by annealing. (a) XPS spectrum of oxygen 1s and (b) DSC heat flow versus temperature for HEO, HPT-processed HEO/CuO and annealed HEO/CuO.

The absorption of light for the three catalysts is illustrated in Fig. 6a using UV-Vis absorption spectroscopy. The HEO shows the presence of an absorption edge at 360 nm, relating to a direct bandgap of 2.6 eV based on the Kubelka-Munk method, as illustrated in Fig. 6b. It should be noted that the estimation of bandgap by the Kubelka-Munk method contains some errors and uncertainties, but this range was reported to be about ± 0.1 eV in the literature [58,59]. The heterojunction HEO/CuO composites after HPT processing exhibit better light absorption than the HEO in UV, visible and infrared regions, but its light absorption slightly decreases after annealing. Although it is hard to estimate the bandgap of the HPT-processed HEO/CuO due to significant background light absorption due to crystal lattice defects, the bandgap for the annealed HEO/CuO is estimated as 2.3 eV in Fig. 6b. XPS spectra were used to calculate the VBM of the catalysts, as shown in Fig. 6c. If the VBM vs. NHE (normal hydrogen electrode) is assumed approximately equal to the VBM vs. Fermi level, the VBM values for the HEO, the HPT-processed HEO/CuO and the annealed HEO/CuO can be estimated from Fig. 6c as 2.0, 1.7 and 1.7 eV vs. NHE, respectively. The VBM level for heterojunctions slightly decreases with the addition of CuO. This is caused by the variations of the coordination environment of $Cu^{2+}$ cations interacting with the HEO, generating occupied defect-like electronic states, and possibly leading to a band shift when compared to the corresponding HEO [60]. Due to the dark color of CuO, its band structure cannot be determined using the facilities available in the authors' laboratory, but the data from the literature can be used for this oxide: bandgap 1.8 eV, VBM 0.7 eV vs. NHE and CBM -1.1 eV vs. NHE [43,61,62]. A summary of the band structure of the annealed HEO/CuO catalyst in Fig. 6d shows the characteristic band structure of a type II heterojunction by assuming that the band structures of the HEO and CuO are not affected by the HPT and annealing processes (due to no changes in crystal structure). In this heterostructured composite, holes are expected to move from the VB of the HEO to the VB of CuO, while the electrons move from the CB of CuO to the CB of HEO [17]. Examining the band structure of the annealed HEO/CuO confirms its compatibility with the energy levels required for the water splitting and $CO_2$ conversion reactions [63,64].

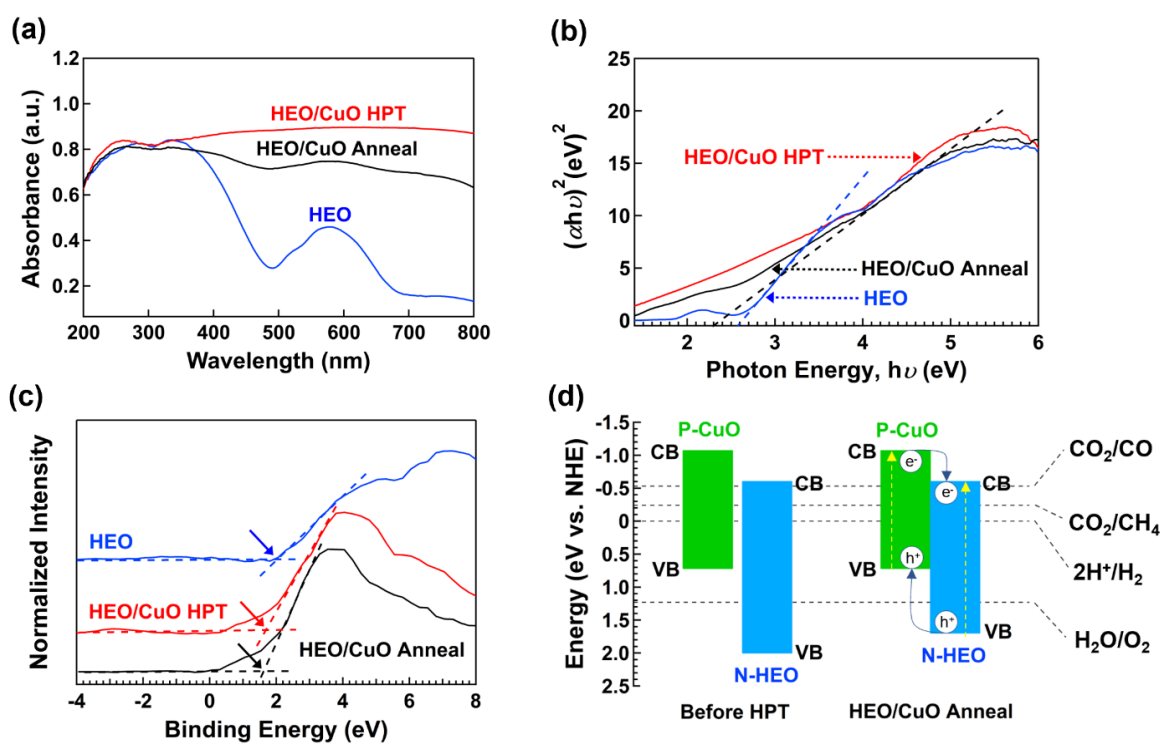

Fig. 6. Enhanced light absorption in heterostructured composite of high-entropy oxide. (a) UV-Vis absorption spectra, (b) direct bandgap calculated using Kubelka-Munk approach ($\alpha$: absorption coefficient, h: Planck's constant, $v$: light frequency), (c) XPS spectra to examine VBM and (d) band structures and potentials of different photocatalytic reactions for HEO, HPT-processed HEO/CuO and annealed HEO/CuO.

Charge carrier dynamics, examined by (a) steady-state photoluminescence, (b) time-resolved photoluminescence and (c) photocurrent measurement, are shown in Fig. 7. As illustrated in Fig. 7a, a photoluminescence peak appears at around 600 nm for all three catalysts, but its intensity decreases with the generation of a heterojunction between the HEO and CuO. Such a decrease indicates that the radiative recombination of electrons and holes is suppressed by the P/N heterojunction, although defects and exciton-phonon interactions can also partly affect photoluminescence [65,66]. Time-resolved photoluminescence spectra shown in Fig. 7b indicate that the carrier lifetime is 8.5 ns, 31.4 ns and 7.1 ns for the HEO, the HPT-processed HEO/CuO and the annealed HEO/CuO, respectively. These values suggest that the annealed HEO/CuO exhibits the fastest charge separation and mobility for the transfer of electrons to the surface due to the contribution of heterojunctions; however, the HPT-processed HEO/CuO exhibits the slowest mobility, possibly because of oxygen vacancies. Fig. 7c depicts the light response to form the photocurrent of the three catalysts. Due to the differences in interaction and bonding features between the catalysts with different particle sizes and the FTO glass, quantitatively comparing photocurrent intensity is quite challenging. However, a few key points can be concluded based on the shape of the photocurrent spectra and qualitative analyses. First, by considering the configuration of the photocurrent system used by the authors, the shape of the photocurrent spectrum for the HEO indicates the characteristic structure of an N-type semiconductor [67,68]. Second, for the HEO, photocurrent increases sharply with turning on the light, and it decreases sharply with turning off the light, indicating rather fast recombination of charge carriers. Third, the photocurrent intensity gradually increases for the HPT-processed and annealed heterostructured composites by turning on the light and gradually decreases by turning off the light. This behavior suggests a slower recombination of charge carriers in heterostructured composites. These results indicate that the HEO/CuO composites have the highest potential to act as a photocatalyst [67,68].

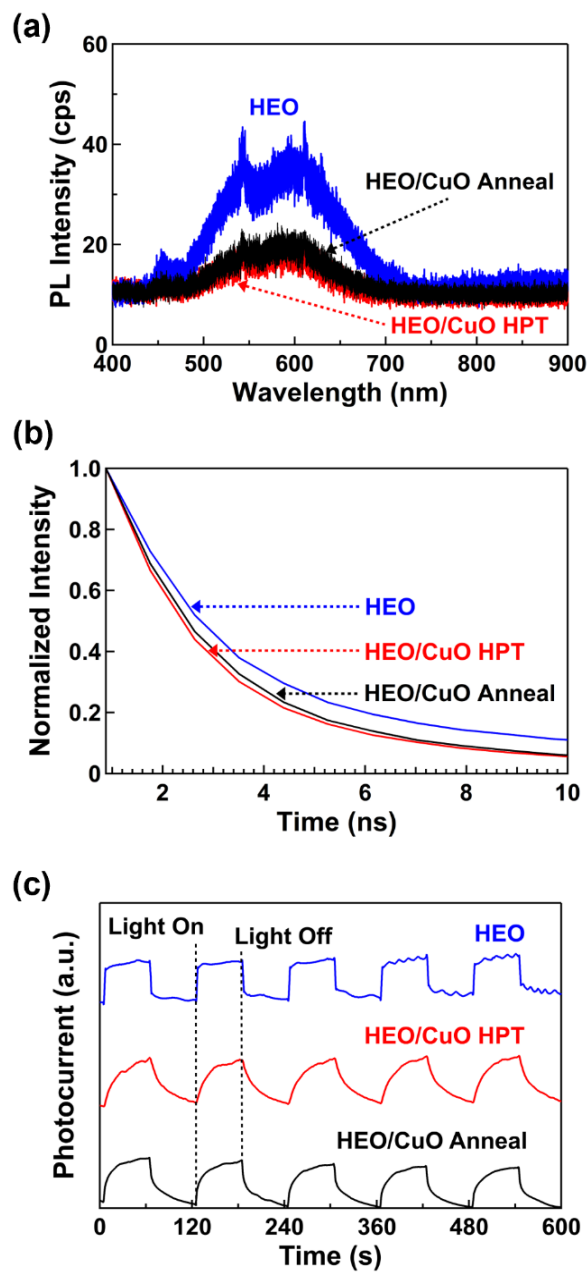

Fig. 7. Suppression of charge carrier recombination by making heterostructured composite from high-entropy oxide. (a) Steady-state photoluminescence, (b) time-resolved photoluminescence and (c) photocurrent spectra for HEO, HPT-processed HEO/CuO and annealed HEO/CuO (ON time: 60s, OFF time: 60s).

### 3.2. Photocatalytic activity

Fig. 8 illustrates the results of two experiments: first, (a) $CH_4$ and (b) CO formation from $CO_2$ photoconversion, and second, (c) $H_2$ evolution from water splitting. For $CO_2$ conversion, the reaction was first conducted for 8 h to reach a steady chemical state, and after that, measurements were recorded every 1 h for a total irradiation time of 24 h. For $H_2$ production, the measurements were conducted for 4 repetitive cycles with a total irradiation time of 12 h. For $CH_4$ production, the annealed HEO/CuO catalyst exhibits the highest efficiency: 1.16, 1.37 and 2.40 µmol/g.h for the HEO, the HPT-processed HEO/CuO and the annealed HEO/CuO, respectively. These findings suggest that the efficient introduction of P/N heterojunction can enhance $CH_4$ production, which is a difficult reaction due to the involvement of eight electrons [69]. Fig. 8b illustrates the production of CO as an intermediate product in $CH_4$ production, using the three catalysts. There is a decrease in CO concentration for the HPT-processed and annealed HEO/CuO catalysts compared to that of the HEO, possibly due to their higher selectivity for $CH_4$ production. The selectivity for $CH_4$ formation can be calculated from the rate of $CH_4$ production, $r_{CH4}$, and the rate of CO production, $r_{CO}$, using the following equation [39].

Methanation Selectivity = 100 x (8$r_{CH4}$) / (2$r_{CO}$ + 8$r_{CH4}$)  (3)

In this equation, the numbers 8 and 2 represent the number of electrons needed for $CH_4$ and CO production, respectively. By using the data after 24 h of $CO_2$ conversion, it concludes that the selectivity for $CH_4$ production is 51%, 60% and 72% for the HEO, the HPT-processed HEO/CuO and the annealed HEO/CuO, respectively. Fig. 8c depicts $H_2$ evolution from the water-splitting process on the catalysts. It observes that the HEO generates an $H_2$ concentration of up to 1.03 mmol/g within 3 h, while the HPT-processed HEO/CuO achieved 0.93 mmol/g after 3 h. The annealed HEO/CuO demonstrated significant $H_2$ production, reaching 2.12 mmol/g after 3 h of irradiation. All samples exhibit excellent cycling stability with no reduction in the production rate. Additionally, XRD profiles of the photocatalysts prior and after the photocatalytic experiments, shown in Fig. 8d, confirm that prolonged irradiation conditions and photocatalytic reaction did not affect the structure of the catalysts. The findings in Fig. 8 highlight that the annealed HEO/CuO has superior activity and stability in both methanation and water splitting due to P/N heterojunctions and optimized vacancy concentration [17,18,43,45,67], but its long-term performance over days and weeks remains a topic of future studies.

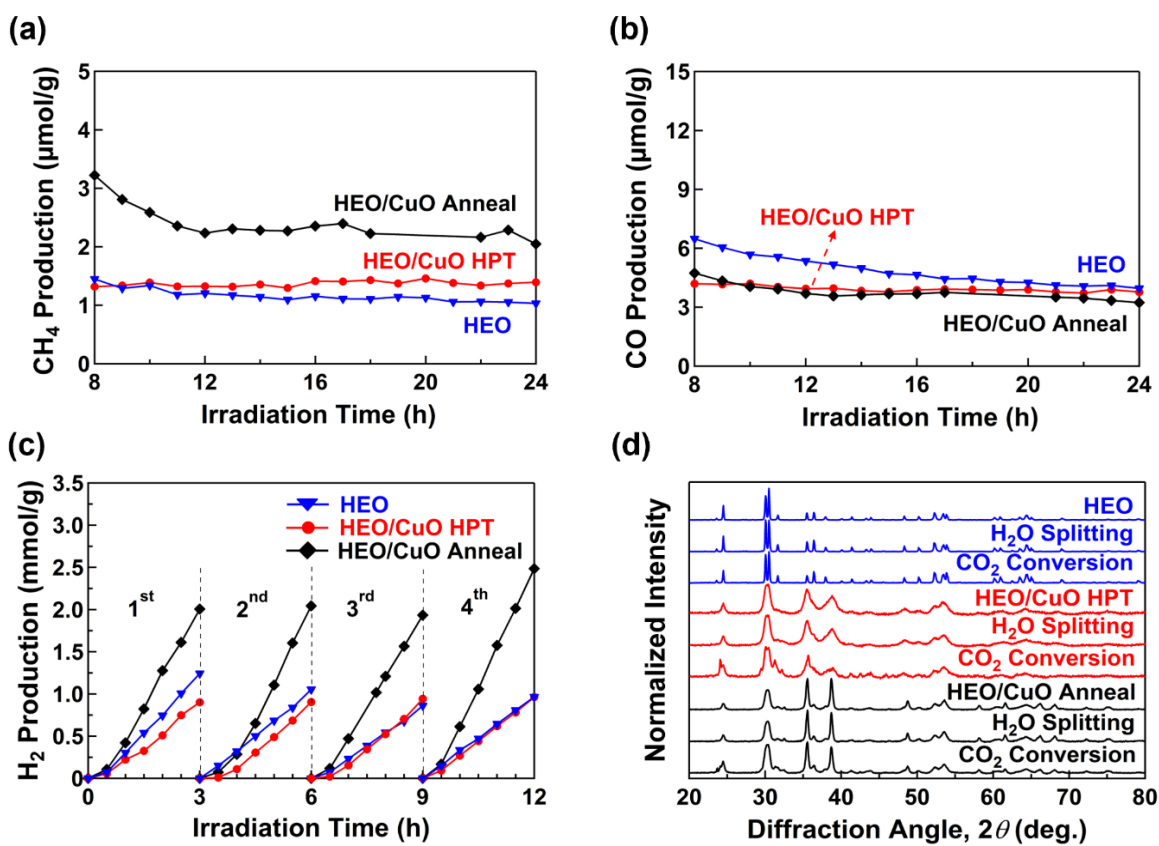

Fig. 8. Enhanced photocatalytic methane and hydrogen production by including heterojunctions in high-entropy oxide containing atomic $d^0/d^{10}$ electronic junctions. (a) $CH_4$ and (b) CO generation from $CO_2$ photoconversion, (c) $H_2$ evolution from water splitting for four cycles, and (d) XRD profiles prior and after photocatalysis for HEO, HPT-processed HEO/CuO and annealed HEO/CuO.

## 4. Discussion

In this study, a strategy for producing active photocatalysts for green $CH_4$ and $H_2$ formation was introduced by adding P/N heterojunctions in an HEO with $d^0/d^{10}$ atomic junctions. Although the significance of mixed $d^0$ and $d^{10}$ electronic configuration cations, as a mixture of electron donors and acceptors at the atomic scale, was discussed in recent publications [39,40], three essential questions should still be discussed in this article. (1) How does introducing a P-type catalyst CuO into an N-type HEO structure affect the optical properties? (2) What is the necessity of removing oxygen vacancies from the P/N heterostructured composite? (3) How does the activity of the P/N heterostructured composite compare with earlier studies?

To answer the first question, we need to compare the configurations of the two catalysts, the HEO and the annealed HEO/CuO. The HEO has good light absorption capability mainly within the UV range and partly in the visible light range. Besides, the HEO has a relatively low bandgap (2.6 eV) and a good band structure suitable for photocatalytic processes (Fig. 6d). However, the annealed heterostructured composite shows significantly better light absorption. This is because when a proper quantity of CuO is added to the HEO structure, it can create a charge compensation trend, which extends the light absorption even into the infrared range [70]. The CBM of the heterojunction is similar to that of the HEO, but the band gap and positions of VBM decrease in the presence of CuO, providing higher overpotentials for both $CH_4$ and $H_2$ production [70]. Moreover, it observes that the HEO shows a sudden increase and decrease in photocurrent intensity, while the annealed heterostructured composite exhibits gradual changes due to lower radiative recombination. The PL spectra (Fig. 7a) and time-resolved PL (Fig. 7b) also confirm that the recombination between electrons and holes in the heterostructured catalyst is smaller compared to the HEO, and the charge carrier lifetime is higher in the presence of P/N heterojunctions due to charge carrier movements between the two phases [17,18,43,45,67].

To address the second question, it is noted that the HPT process induces significant lattice distortions and creates a large density of oxygen vacancies [25]. Analyses in Fig. 5 confirm the existence of such vacancies in the HPT-treated HEO/CuO. While oxygen vacancies on the surface can function as sites for photocatalytic reactions [24,47,71–73], their excessive presence in the bulk hinders the mobility of charge carriers (electrons or holes) within the crystal lattice [49]. This reduced mobility lowers the efficiency of the photocatalytic reactions. Based on the results of $CO_2$ conversion and water splitting, the HPT-processed HEO/CuO shows almost similar activity to the HEO despite the generation of P/N junctions. Implementing a post-HPT annealing process reduces the oxygen vacancy concentration, thereby enhancing the mobility of charge carriers, as shown in Fig. 7b employing time-resolved photoluminescence spectroscopy, and improving the overall photocatalytic performance of the heterojunction [49]. It should be noted that there are three primary pathways for $CO_2$ conversion: carbene, formaldehyde and glyoxal [8,69,74]. Based on Fig. 8, the products of $CO_2$ conversion include $CH_4$ and CO, while no HCOOH is detected. This indicates that the carbene pathway is likely the dominant pathway, as opposed to the formaldehyde and glyoxal pathways. The presence of excess oxygen vacancies appears to deteriorate the activity in the carbene mechanism.

To address the third question, a comparison was made, as shown in Table 1, to evaluate the performance of the annealed heterostructured composite for methanation and hydrogen production with respect to the reported data in the literature, including some P/N, Z-scheme and S-scheme heterojunctions [18,75–92]. In Table 1, based on the particle size measurement by SEM, the surface area of the annealed HEO/CuO is estimated as 0.055 m$^2$/g. Such a small surface area, which could not be quantified by nitrogen gas absorption and the Brunauer-Emmett-Teller (BET) method, is due to the synthesis by HPT (high pressure and high strain lead to undesirable consolidation of powders [93]). Although the experimental conditions are different in different studies, and the surface areas of the catalysts are quite small after synthesis by HPT [38], the HEO-based P/N heterostructured composite is still one of the highly efficient heterojunctions for photocatalytic $CH_4$ production as well as $H_2$ production. The activity of the photocatalyst is particularly high when the activities are compared per surface area of the catalyst in Table 1. The findings of the present work confirm the high potential of HEOs with atomic $d^0/d^{10}$ electronic junctions and P/N heterojunctions as new photocatalysts for green $CH_4$ and $H_2$ production, although their activity should be quantified by quantum efficiency or turnover frequency measurements in future studies.

It should be noted that while the mechanism of high photocatalytic activity of P/N heterojunctions in conventional photocatalysts has been widely investigated by considering the electron-hole separation (Fig. 6d) [16–20], future experimental and theoretical studies (e.g., in situ irradiation XPS, electrochemical impedance spectroscopy, density functional theory (DFT) calculations, etc.) should clarify the mechanism, kinetics and pathway of charge carrier separation in high-entropy P/N heterojunctions. Moreover, although HPT produces a fast and reproducible synthesis route for (i) photocatalysts and high-entropy catalysts [23], (ii) new high-entropy materials [94] and (iii) dual-phase heterostructures/composites [95], future studies should focus on producing P/N heterojunctions with large-scale and high specific surface area.

Table 1. Comparison of photocatalytic performance across studies of heterostructured P/N, Z-scheme and S-scheme composites for hydrogen and methane formation in comparison with high-entropy oxide with $d^0/d^{10}$ electronic junctions and P/N heterojunctions.

| Material | Heterojunction type | Surface area ($m^2/g$) | Photocatalytic activity | | Ref. |
|---|---|---|---|---|---|
| | | | CH$_4$ production | H$_2$ production | |
| P-CuO/N-HEO | P/N type (type II) | 0.055 | 2.40 µmol/h.g (43.64 µmol/h.m$^2$) | 0.71 mmol/h.g (12.91 mmol/h.m$^2$) | This study |
| g-C$_3$N$_4$/BiVO$_4$ | Z-scheme (type II) | 37.30 | 4.57 µmol/h.g (0.12 µmol/h.m$^2$) | - | [18] |
| g-C$_3$N$_4$/Nb$_2$O$_5$ | Type II | 45.08 | 16.07 µmol/h.g (0.36 µmol/h.m$^2$) | - | [75] |
| CsPbBr$_3$/BiOCl | S-scheme (type II) | - | 3.47 µmol/h.g | - | [76] |
| WO$_3$/In$_2$O$_3$ | S-scheme (type II) | - | 5.4 µmol/h.g | - | [77] |
| Fe$_2$O$_3$/Graphene/Bi$_2$O$_2$S | Z-scheme (type II) | 17.24 | 4.27 µmol/h.g | - | [78] |
| g-C$_3$N$_4$/InVO$_4$ | Type II | - | - | 0.21 mmol/h.g | [79] |
| Ta$_2$O$_5$/g-C$_3$N$_4$ | Type II | - | - | 0.04 mmol/h.g | [80] |
| P-Cu$_3$Mo$_2$O$_9$/N-TiO$_2$ | P/N type (type II) | 107 | - | 3.40 mmol/h.g (0.03 mmol/h.m$^2$) | [81] |
| NiS/C$_3$N$_4$ | - | - | - | 0.48 mmol/h.g | [82] |
| P-NiO/Ni$_2$P/N-CN | P/N type (type II) | 56.3 | - | 0.50 mmol/h.g (0.01 mmol/h.m$^2$) | [83] |
| P-Bi$_2$O$_3$/N-MoS$_2$ | P/N type (type II) | 159.3 | - | 10 µmol/h.g (0.06 µmol/h.m$^2$) | [84] |
| P-MnS/N-CdS | P/N type (type II) | - | - | 5.92 mmol/h.g | [85] |
| P-CuO/N-In$_2$O$_3$ | P/N type (type II) | - | - | 0.60 mmol/h.g | [86] |
| P-Cu$_2$O/N-NiFe$_2$O$_4$ | P/N type (type II) | 11.74 | - | 4.00 µmol/h.g (0.34 µmol/h.m$^2$) | [87] |
| P-CuInS$_2$/N-ZnInS$_2$ | P/N type (type II) | - | - | 0.29 mmol/h.g | [88] |
| g-C$_3$N$_4$/WO$_3$ | Z-scheme (type II) | 34 | - | 3.12 mmol/h.g (0.09 mmol/h.m$^2$) | [89] |
| TiO$_2$/CdS | Z-scheme (type II) | - | 11.9 mmol/h.m$^2$ | - | [90] |
| g-C$_3$N$_4$/g-C$_3$N$_4$ | S-scheme (type II) | 41 | - | 0.598 mmol/g.h (0.015 mmol/h.m$^2$) | [91] |
| TiO$_2$/CdS | S-scheme (type II) | 37 | - | 2.320 mmol/h.g (0.063 mmol/h.m$^2$) | [92] |

## 5. Conclusions

In this study, a heterostructured composite integrated from P-type CuO and N-type high-entropy oxide containing atomic $d^0/d^{10}$ electronic junctions was synthesized and applied for CH$_4$ and H$_2$ production from CO$_2$ conversion and water splitting, respectively. This combination enhances the optical properties of the material, such as expanding the light absorption range, reducing the recombination rate of carriers and boosting charge carrier mobility. The performance of the heterojunction composite depends on oxygen vacancies, while partial annihilation of vacancies improves both $CH_4$ and $H_2$ production rates. These results open up new potential for high-entropy oxides with atomic $d^0/d^{10}$ electronic junctions and P/N heterojunctions for use in photocatalytic processes. The concept of this study can be extended to fabricate a wide range of new catalysts for not only water splitting and $CO_2$ conversion but also other reactions, such as water treatment, plastic waste degradation, ammonia production, etc.

**CRediT authorship contribution statement**

All authors contributed to conceptualization, investigation, methodology, validation, and writing - review & editing.

**Declaration of competing interest**

The authors declare no conflict of interest.

**Acknowledgments**

The author H.T.N.H. is grateful to the Yoshida Scholarship Foundation (YSF) for a Ph.D. scholarship. The current research is funded partly by Mitsui Chemicals, Inc., Japan, partly by a Grant-in-Aid from the Japan Society for the Promotion of Science (JP22K18737), and partly by the ASPIRE project of the Japan Science and Technology Agency (JST) (JPMJAP2332).